\documentclass[12pt]{article}
\usepackage{geometry}
\usepackage[round]{natbib}
\usepackage{graphicx}
\usepackage{amsmath}
\usepackage{amssymb}
\usepackage[utf8]{inputenx}
\usepackage{lineno}
\usepackage{color}
\usepackage[english]{babel}

\begin{document}

\author{J. A. Jiménez Madrid$^{a}$, A. García-Olivares$^{a}$, J. Ballabrera Poy$^{a}$ \\ and \\ E. García-Ladona$^{a}$ \\
\\ 
\small\;$^{a}$ Instituto de Ciencias 
del Mar (ICM-CSIC) \\
\small Passeig Marítim de la Barceloneta, 37-49. E-08003 Barcelona (Spain)}
\normalsize
%
\title{Managing large oil Spills in the Mediterranean}

\maketitle
\begin{abstract}
For the first time a statistical analysis of oil spill beaching is applied to the whole Mediterranean Sea. A series of probability maps of beaching in case of an oil spill incident are proposed as a complementary tool to vulnerability analysis and risk assessment in the whole basin. As a first approach a set of spill source points are selected along the main paths of tankers and a few points of special interest related with hot spot areas or oil platforms. Probability of beaching on coastal segments are obtained for 3 types of oil characterised by medium to highly persistence in water. The approach is based on Lagrangian simulations using particles as a proxy of oil spills evolving according the environmental conditions provided by a hincast model of the Mediterranean circulation.
\end{abstract}


\newcommand{\figura}[2][]{\includegraphics[#1]{#2}}
\newcommand{\rojo}[1]{\textcolor{red}{#1}} 
\renewcommand{\baselinestretch}{2}

\normalsize

\section{Introduction}

Oil spill pollution enters the marine environment through two pathways. The most important route is direct pollution associated to global maritime trade where oil and related substances are released in ship accidents (e.g. Exxon Valdez, Prestige,\ldots), from illegal procedures (bilge waters release) and associated to offshore industry, typically oil-gas drilling platforms (i.e. Deepwater Horizon). A second but less frequent way is from land to sea, associated to releases in coastal areas coming from industrial accidents or collateral war-related damages (e.g. War of Gulf in 1991, Lebanon spill in 2006). The major number of large oil spills events is associated to ship accidents, mostly damaged oil tankers but also fuel leaking from non-tanker ships incidents. According to recent statistics from the International Tanker Owners Pollution Federation (ITOPF), the cumulative data of oil spills from oil tankers in the last forty years show a decreasing tendency of both the total number of oil spills and the total amount of oil spilt \cite*[see figure 2 in][]{itoff2013}. This evolution is in part explained by technological improvements in ship design and navigation security but also as a consequence of more severe regulations adopted through international agreements \citep{solas1974,marpol1978,opa1990,ismcode1993}. In the 1990's, 73\% of the amount of spills over 7 t was produced by just 10 incidents. At the beginning of this century, with approximately half of oil spills compared to the preceding decade, 47\% of this amount was spilt in just 2 incidents. So large spills still constitute a major fraction of the total amount of oil spilt in marine water.

When one of such incidents occurs the consequences linked to environmental pollution, cleaning operations, and economic impacts are enormous; but what makes the difference among them are the particular circumstances around the event. According to \citet{whitemolloy2003} there is a common consensus on the main technical factors influencing the costs of spills. These factors can be enumerated in the following way: (i) The persistence of the oil in the environment; (ii) weather and sea conditions; (ii) amount spilled and rate of spillage; (iv) effectiveness of clean-up, and (iv) geographical, biological and economic characteristics of the affected area. 

The cumulative number of ITOPF attendances to oil tankers incidents since 1970 shows the European and Asiatic coastal regions as the most affected areas with a major number of spill events. The Mediterranean Sea, still the shortest route from Asia to Europe, is another example of such concentration of incidents. About 1/6 of the global maritime traffic and 1/3 of the global seaborne oil (almost 8 million oil barrels per day) is carried through the Mediterranean Sea which represents only 0.8\% of the ocean surface. In a report by the Regional Marine Pollution Emergency Response Centre for the Mediterranean Sea \citep[]{rempec2011}, it has been estimated a total spillage of around 310 000 t of oil since 1977 and at least 120 000 t of other noxious substances since 1988. There has been a total of 659 accidents with 545 involving oil substances \citep[see page 5 in][]{rempec2011}. The updated data, available since 1977, also exhibit a decreasing trend since the 90's, both for the quantity of oil and other hazardous and noxious substances and only disrupted by the major accidents of MT-HAVEN and Lebanon spills. The MT-HAVEN accident has been by far the largest in the Mediterranean, not only because the spilled quantity was quite huge (about 145 000 t, among the top ten biggest spills from tankers) but also due to the environmental damage. The accident caused a serious pollution of sea waters, seabed and along the Ligurian coast, from Genoa to Savona, producing a shipwreck that still today constitutes a potential source of pollution.

Besides the amount released in the MT-HAVEN incident was comparable, and even greater, than other devastating accidents over the world ocean (e.g. Deep Water Horizon, Prestige, Erika, Exxon Valdez) the environmental impact was of relative lesser extent because the circumstances were quite distinct. The proximity to the coast, the rapid response and the environmental conditions were elements that limited the damage if it is compared, for example, with respect to the Prestige, the Erika, or the Exxon Valdez, in which more than 1000 km, 400 km and 800 km of coast were respectively affected. Thus the environmental impact produced by oil spills is a complex combination of several individual factors and circumstances that renders difficult to totally prevent it.

In case of an oil spill accident, the most pressing issue is to know whether the oil slick will go onto a coastal area in the following days. Even for state of the art oil spills forecasting models, the prediction horizon is about 5-7 days because this is in fact imposed by the horizon of a meteorological forecasting. If a 7-day forecast predicts that slicks still will have not reached any shoreline, the most exposed coastal segments to the long term slick drift remain uncertain. To figure out which is the most exposed coastline, we suggest to take into account statistical information of coastal segments exposed to spills (in the scale of months) coming from the most probable potential sources, such as tankers paths, oil-gas drilling platforms, etc. 

The short term response can be assessed through combined meteorological-oceanographic forecasting systems and oil spill forecast models. The long term assessment can be based on statistical analysis of risk of oil-spill arrival to sensitive areas based on past environmental conditions, namely winds and currents. This procedure was applied by the US Department of Interior to evaluate environmental consequences prior to lease sales or approval of industry's plans \citep{guillenetal2004,priceetal2004,johnsonetal2005}. A similar approach has been used to elaborate an oil spill risk assessment for the Cantabrian coast, the northern coast of Spain in the Bay of Biscay \citep{juanesetal2007,castanedoetal2008}. \citet{garciaetal2011} reanalysed the risk assessment procedure applied to the Prestige case in order to evaluate the costs of solutions adopted by policy-makers, in front of possible alternatives based on prior knowledge about the statistics of beaching. 

In this paper we present the methodology and first results of beaching probability of spills for the whole Mediterranean Sea. In MEDESS-4MS project (http://www.medess4ms.eu/), a multimodel oil spill forecasting system has been developed to cope with the short time forecasting needs to face pollution events in the Mediterranean Sea. As part of the vulnerability analysis and risk assessment procedures, we have undertaken a statistical approach of spills trajectories from available hindcast of the Mediterranean circulation. In section~\ref{sec:methodology} we describe in detail the methodology, the data source and the oil-spill model to compute the beaching probability. In section~\ref{sec:results} we illustrate the results with some examples and we end by summarising the main conclusions.

\section{Material and Methods}

\label{sec:methodology}
Forecasting the long-term behaviour of an oil spill is not an easy task, since ocean and atmosphere are highly turbulent fluids and the spill dispersion is very influenced by the changing structures of these geophysical flows. 
To better understand the long-term behaviour of an oil spill and the evolution of structures in such flows we propose to make use of statistical analyses of many Lagrangian simulations based on hindcasts of atmospheric and oceanic flow fields observed in past years. We will carry this statistical analysis for a sets of spills locations within the Mediterranean basin.

Our approach focus on identifying the most probable sources of danger of oil spills in the Mediterranean Sea, mainly composed of downloading infrastructures, oil and gas platforms, pipelines, bunkering regions and tankers paths. However, given the data availability of many of them, we have restricted the analysis to the main routes of tankers within the basin. These have been reported in the geographical information system (GIS) on marine trade in the Mediterranean elaborated during the SAFEMED project (see http://www.safemedgis.org/). Figure~\ref{fig:spillpoints} depicts a set of points along main tanker trajectories every 70-100 km. The major traffic activity is concentrated along the route from Strait of Gibraltar to Suez Channel as being the natural path to the north of Europe. Additionally several Mediterranean harbours with downloading facilities as Genoa, Marseille, Tarragona, Arzew, together with the trade line linking the Black Sea define other secondary routes in terms of oil trade activity considered as potential sources for incidents involving spills. 

\subsection{Hindcast data}

Given the point sources, we follow a Lagrangian approach where spills are assumed as parcels that are primarily dragged by winds and advected and diffused by ocean currents. Thus the time evolution of spills can be computed using essentially winds and currents provided by hindcast simulations. For this study, we use NEMO-MED12 model outputs for the ocean currents \citep{JGRC:JGRC12389}, a regional configuration of the NEMO model \citep{madec2008}, and winds for ARPERA \citep{GRL:GRL24117} atmospheric model, which is a dynamical downscaling of ECMWF products. The studied period covers the years 1998 to 2007. The NEMO-MED12 model system has a grid configuration stretched in latitude of 1/12º of resolution, which corresponds to a cell range of 6-8 km; so at the limit of eddy resolving and eddy permitting according to the characteristic Rossby radius of deformation for the Mediterranean (about 15 km). It has a variable vertical resolution with 50 z-levels being of about 1 m at the surface, 20 m at 100 m depth ($23^{rd}$ z-level), and 460 m at the bottom. This ocean model has been forced by the daily ARPERA fluxes and wind stress, at 50 km resolution \citep{JGRC:JGRC12389}. Figure~\ref{fig:velocityfield} shows, as an example, a map of the oceanic surface velocity field from this hindcast. The major well known features  (the Algerian current, the Mid-Ionian jet, the Liguro-provezal current, etc) of the Mediterranean surface circulation appear in agreement with the classical circulation scheme \citep[e.g.][]{millot2005}.

\subsection{Oil Spill model}

In order to have affordable computation time, the evolution of oil spills is done according to a simplified model. The oil spill model accounts for the pure advection by mean currents and wind drag over as the main relevant elements determining the oil spill evolution. The Lagrangian model is then formulated in the following way: at each point, a number $N$ of particles are released every day. Each particle is moved according to the following differential equation:
\begin{equation}
\frac{d\mathbf{x^i}}{dt} = \mathbf{u^i_C} + \alpha\; \mathbf{u^i_W} \ , \label{eq:adv}
\end{equation}
where $\mathbf{x^i}$ is the vector position of the particle labelled $i$. This particle is advected by the water current $\mathbf{u^i_C}$, and the wind action $\mathbf{u^i_W}$. The coefficient $\alpha$ is related with the wind drag and $\mathbf{u^i_W}$ is the wind velocity at $10\;m$. Because wind forcing in the hindcast fields is provided in terms of wind stress at the surface, we need the $\mathbf{u^i_W}$ at standard $10\;m$ level as a function of the stress at the surface. To do that we make use of the boundary layer logarithmic law near the air-water surface \citep{batchelor1967}: 

\begin{equation}
U(z) = \left(\frac{u_\ast}{k}\right) \ln\left(\frac{z}{z_0}\right), \label{eq:ln}
\end{equation}
where $U(z)$ is the mean wind at any level $z$, $k$ is the von Kármán constant ($k\simeq 0.4$), $u_\ast$ is the characteristic velocity associated to the intensity of turbulent fluctuations and $z_0$ is the roughness length. If we consider that,
\begin{equation}
\tau_s = \rho_a\;u_{\ast}^2 \ ,
\end{equation}
where $\rho_a$ is the air density and the scaling law proposed by \citet{charnock1955}, we have
\begin{equation}
 z_0 = \frac{\beta u_{\ast}^2}{g} = \frac{\beta \tau_s}{\rho_a g} \ ,
\end{equation}
where $\beta = 0.016$ \citep[see page 381 in][]{stull} for the sea. Now, we can express the wind drag term in equation \ref{eq:adv} as a function of the surface stress by substituting the roughness length in equation~\ref{eq:ln} and setting $z=10\;m$. By applying the previous reasoning to the meridional and zonal component of velocity, we get
\begin{equation}
\frac{d\mathbf{x^i}}{dt} = \mathbf{u^i_C} + \frac{\alpha}{\sqrt{\rho_a} k} \left(\sqrt{\tau_{s_U}}\; \ln\left(\frac{10\;\rho_a g}{\beta \;\tau_{s_U}}\right)\; , \; \sqrt{\tau_{s_V}}\; \ln\left(\frac{10\;\rho_a g}{\beta \;\tau_{s_V}}\right)\right)\ ,\label{eq:final}
\end{equation}
where $\tau_{s_U},\; \tau_{s_V}$ are the zonal and meridional (respectively) wind stresses at the surface, and $\alpha \simeq 0.03$ according to \citet{Spaulding1988}.

At this point it is opportune to mention some details about the numerical integration of equation~\ref{eq:final}. For advecting particles, classical Runge-Kutta methods are advised \citep{pressetal1993}. The order $n$ for the Runge-Kutta method means that the error per step is on the order of $O(h^{n+1}),$ and the total accumulated error has order $O(h^4),$ where $h$ is the step-size. Most of models implement a fourth/fifth-order method so here we use a fifth-order method. As a tricky peculiarity, computed fields are usually given by a grid of functional values in a discrete set of points, so we need to interpolate field values at some untabulated point within the grid cell. Common approaches use bilinear interpolation which is the simplest interpolation method and often enough to study the general behaviour of a field. As the interpolated point changes from cell to cell the interpolating function varies continuously, however the gradient of the interpolating function changes discontinuously at the cell boundaries. These discontinuities could produce numerical artifacts and a weird behaviour in the model results for a long time integration as it is here the case.

So higher-order methods to increase accuracy for the interpolating function trying to fix up the continuity of the gradient are recommended \citep[see section 3.6 in][]{pressetal1993}. For these reasons \emph{bicubic interpolation} is suggested to satisfy these properties. Bicubic interpolation requires to specify function values, gradients and the cross derivative at each grid point. Then the interpolating function satisfies the following properties: (i) the values of the function and the specified derivatives are reproduced exactly at the grid points, and (ii) the values of the function and the specified derivatives change continuously as the interpolating point crosses from cell to cell. It is important to notice that bicubic interpolating equations do not require to specify the extra derivatives correctly as the smoothness properties are forced by construction of the method. But the better you provide the required values, the more accurate the interpolation will be (the interpolating function will always be smooth). It is easy to use centred differencing to compute all the values for the derivatives required by this interpolating scheme.

Also, and for the same reasons, it is suggested to go beyond the lineal interpolation for the time coordinate. There are several choices but the most appropriate combination in terms of satisfying good mathematical properties and accuracy for spatial-time interpolation is to use bicubic spatial interpolation in space and third order Lagrange polynomials in time. There are interpolating methods in time with higher accuracy but their computational cost is huge (as can be seen in the detailed comparison given in \citealt{mancho2006}). So the choice here gives accurate results at a moderate computational cost.

We end this subsection with further considerations about the oil spill model and the beaching criteria. First, notice that any oil spill model should consider complex oil weathering processes \citep[e.g.][]{coppinietal2011, dedominicis2013a, dedominicis2013b}. Instead of accounting with all detailed weathering processes, we propose a simplified way to approach them by distinguishing between 4 oil types according to their persistence in water \citep*[]{itoff2014}. The persistence curves for the four oil groups are shown in figure~\ref{fig:persistence}. For the oil of type 1 the evaporation is very fast, so such oil fractions will disappear before reaching the coast in most of the cases. Then, only risk maps for oil types 2, 3 and 4 have been considered. Second, the statistical analysis of beaching is done by a regular release of virtual particles at the point source. A total number of 185 points along tanker paths has been retained as hypothetical potential point sources of oil spills. We have chosen to release $N=90$ particles every day at each point source during the whole period of 9 years; which is equivalent to a total amount of roughly 300 000 tracking trajectories per point source. To illustrate the results, we have restricted the study to some specific cases distributed along the Western Mediterranean basin (points along tankers paths), an offshore oil platform and one characteristic point located in the Gibraltar Strait area. Besides, for the sake of simplicity we are using here only risk maps with constant persistence. Risk maps with constant persistence may also serve as a proxy of other kind of pollutants, as plastics or similar, having very long persistence at water.

The beaching criteria considered in these simulations is that a particle has reached the coast when it moves less than about $500\;m$ during the last day and it is close to the coast (the particle is within a grid cell where at least one vertex is land). The persistence time for each oil type is used by taking into account the time required for the spills to reach the coast. The beaching probability is then computed from the fraction of particles which have reached a particular coastline segment. To illustrate the methodology we have decide to divide arbitrarily the Mediterranean coastline into segments of $50\;km$ length to produce beaching probability maps. One may wonder to what extent a hindcast of 9 years provides enough statistics or even if the hindcast simulation still contains some contamination of the spin-up phase of the simulation. We have computed the kinetic energy of the model for the surface velocity layer and we have compared all the risk maps generated for a given point by chopping the data every 100 days, thus taking into account the evolution from the selected day until the end of the simulation. We have generated 32 risk maps with this procedure (not shown) and the distribution of particles reaching the coast remain almost unaffected between beaching maps. Only risk maps for the last 200 days of the hincast period exhibit some differences because particles do not have enough time to reach the distant coasts. However the time evolution of the kinetic energy suggest that the model simulation reach a reasonable stationary state after the first 200 days of the simulation and this is the period considered in dealing with the beaching probability maps.

\section{Results}

\label{sec:results}

The first point is related with an active oil drilling platform, the Casablanca platform operated by REPSOL S.A. and located on the northeast shelf of the Western Mediterranean Sea, just in front of Ebro's Delta (figure~\ref{fig:casablanca}). In the figure one can see that the coast slightly north and around Barcelona would be affected by an oil spill coming from the platform. Quantitatively speaking if an oil spill of $100\; t$ occurs at the platform, then over $15\;t$ of oil will reach the coast around Barcelona area. But there are other places with non negligible probability (green, yellow and orange colours) which are also affected. Notably the northwestern coast of Mallorca with similar pollution levels as in the Barcelona area. By order of importance, another segments along the coastline of Gulf of Valencia and other areas as the northern coast of the Balearic Islands and the region north of Barcelona would also be affected. To remark that the area around the Ebro's Delta, while having relatively medium to low probability, is a high vulnerable area. The simulations show that approximately $5\;t$ of oil could reach the Ebro's Delta. This value is relatively low but given the high vulnerability, a great ecological and biological impact in this area should be expected. It is worth to notice that some segments of the northern Tunisian coast receive a relatively medium quantity of oil while many other closer areas are practically unaffected.

Another region of high interest is the Gibraltar Strait, one of the hot spots in the Mediterranean in terms of ship traffic and marine security. We have taken a representative point located in the middle of the Strait of Gibraltar (figure~\ref{fig:gibraltar}). In this case we have performed two kinds of simulations, one considering the whole period of time for the hindcast, and the other only considering the subset of particles released during the first fortnight of every September. In the first case (upper plot of figure~\ref{fig:gibraltar}) all the segments in the north and south parts of the Alboran basin receives oil, being the northern coast around Algeciras bay and around Malaga the most relevant. In the south, the segment in the Gulf of Hoceima, at the left side of Cape Tres Forcas in the Morocco coast, is an area particularly impacted. A non negligible fraction of around $2\%-3\%$ (green and yellow vertical lines) reaches the Atlantic side. There appear some differences in the second case (lower plot of figure~\ref{fig:gibraltar}) where the simulation is performed releasing particles only in the first fortnight of every September. The north coast in the region will receive less pollutant, for example around Mazarrón (Murcia) oil would not reach this coast at all (see the white segment in the plot). On the other hand there is an overall increase of oil in almost all the segments along the African coast. Further the number of particles quitting the Mediterranean towards the Atlantic basin also increase. This clearly shows that beaching probability maps may depend in general on the time of the year at which the spill is produced. Probably the differences observed are produced by different seasonal wind and currents regimes.

More examples are presented in figure \ref{fig:barrier}. As expected, any coastal segment relatively close to a point source are in general affected. However, it is remarkable the case of a point in front of Marseille (top-right plot in figure \ref{fig:barrier}). The neighbour coastline is relatively little affected, except some segments inside the Gulf of Lions, while the major impact is on the west coast of Sardinia and some segments of the Tunisia coastline. Interestingly, these are the same segments of the Tunisian coast that receive significant oil coming from other points, including the Casablanca case (see figure~\ref{fig:casablanca}). For the cases of Marseille and Barcelona, the Liguro-Provencal current may transport particles southwards towards the Balearic basin, traversing the Ibiza channel and eventually being reincorporated again to the Algerian current.
 
There are evidences of drifters trajectories from the Surface Velocity Program showing behaviours that can explain these patterns \citep{poulainetal2013}. Some drifters released in the Ibiza channel have been observed to reach the Sardinia coast traversing meridionally the middle part of the Algerian basin between the Balearic shelf and the Sardinia coast. The position of the seasonal thermal front across the Algerian basin may also facilitate the recirculation of particles from the Liguro-Provencal current towards Corsica and Sardinia coast.

Another interesting feature arising from these maps is the presence of some unaffected areas. As it can be seen most of the western basin coastline receives some particles although with low beaching probability (blue colour). Besides, most of coastal segments inside the Gulf of Lions and around Marseille receive a few quantity of oil or nothing at all, as for example the case of Sardinia in the lower-right plot in figure \ref{fig:barrier}. This could be an indication of the presence of some transport barriers in the basin, although this can only be verified with a more intensive quantity of release points.

Finally, in figure~\ref{fig:percentages}, we show the time required by particles to reach the coast from two source points. The first one is located at the Casablanca oil platform, where it can be seen that $50\%$ of particles need more than 43.6 days to reach the coast. This highlights the importance to take into account the long time evolution of an oil spill. In the case the spill is not rapidly or easily controlled in less than a week, then actual forecasting systems will not be able to manage in a proper manner the long term evolution of the spill. In the case of Casablanca oil platform, the first particle reaches the coast in 3.1 days and the last one requires 602.1 days with a mean beaching time for the particles of around 62.5 days. 

The second selected point is located at (17.1667º E, 36.0373º N). According to results (figure~\ref{fig:percentages}) one can observe that the time needed to reach the coast from this location is longer than from Casablanca site. Indeed, $50\%$ of particles need more than 73 days to reach the coast, the first particle arrives to the coast in about 16 days and the last one in about 551 days. The mean time required by the particles to beach is around 86 days. These longer beaching times could be produced by eddies in the region which could trap the particles for long time periods. Also, it is interesting to note that after 28 days $36.12\%$ and $1.25\%$ of particles have arrived to the coast from Casablanca and from (17.1667º E, 36.0373º N) respectively. These results qualitatively agree with the maps of ``coastal approach'' obtained by \citep{JGRC:JGRC10492} (see their figure 10) using a different circulation model. The differences in the exact values found here are probably due to the differences in the beaching criteria as we have considered a better definition of the coastline, so we are tracking particles closer to the coast.
 
\section{Summary}

With the purpose of improving vulnerability analysis and risk assessment procedures, we have undertaken a statistical approach to the beaching of oil spills. We have computed the trajectories followed by virtual particles released from some different positions in the Mediterranean Sea. The statistical analysis is based on available hindcasts of 9 years of the Mediterranean circulation. Due to the amount of data only some examples for the western Mediterranean basin are presented here to illustrate the proposed methodology. The sources of danger are supposed to be points along the main tracks of tankers and close to downloading oil facilities and infrastructures as oil drilling platforms. 185 initial points have been selected which imply a total ensemble of 740 maps. The proposed method is able to predict the percentage of initial spill that will beach at a certain coastline segment of interest. For example, if the oil spill is about 30 000 t and the predicted fraction is $5\%$, that means that the segment of interest is expected to receive 30 $t/km$ of oil. 

The method is also capable of estimate the period during which the slicks are expected to affect a given segment, for instance, we have showed the time required from two locations where the first slick would arrive in 3 and in 16 days, respectively.

These maps will be available for the Decision Support System of the MEDESS-4MS system (and freely available in the web) and are expected to improve the decision-making process in case of an accident in the Mediterranean. This information may be complementary to the short-term forecasts provided by models included in the system. In this way, short and long-term predictions would be available since the very beginning to decision making authorities. The information can also be combined with methodologies estimating the economic impact and the ecological risks of sensible areas (\citet{garciaetal2011}).

The beaching probability maps here computed have some limitations. First of all they rely on a hindcast simulation which may no capture all the variability of the Mediterranean circulation. In particular, several aspects as the spatial resolution and the model ability to deal with the dynamical processes in coastal areas, upon which ultimately depend the beaching of particles, are perhaps the most relevant. In addition the oil transport model used here lacks some other transport processes usually present in short time oil weathering spills models as Stokes drift, stochastic dispersion and fate components. However, these are not limitations of the proposed methodology because it can be easily implemented, but instead, they rely on the computationally effort required to run the Lagrangian evolution of particles. The robustness to such terms are left for future work. 

Furthermore, results may depend whether averages are done on a seasonal or monthly basis instead of global averages much more representative of the intermittent dynamics of the Mediterranean circulation where both forcing and predominant currents can not be strictly defined by mean patterns. At the end, this set of maps should not be considered as a definitive product but a first approach. As soon as new hindcasts of the Mediterranean circulation are improved and becoming more realistic, the maps can be recomputed to increase representativity and therefore to improve the level of safety in the decisions that authorities have to take in the first hours following a large spill accident. At the time of written this paper the Copernicus service has released a new and longer reanalysis of the Mediterranean circulation which now can be used to repeat the analysis and see whether the results here presented are still valid and robust.

In summary estimations of long-term beaching probability through Lagrangian simulations are highly valuable pieces of information in the management of an oil spill emergency. They are produced to help decision making procedures when an emergency occurs, but can also be used to identify stable features such as transport barriers to the long term dispersion. Limitations are not insurmountable and depend basically on computationally resources and can be generalised and extended to additional sources of danger or type of pollutant as for example pollution by plastics.

\paragraph{Acknowledgements:} The authors wish to thanks the support from projects MEDESS-4MS (Mediterranean Decision Support System for Marine Safety, MED 2S-MED11-01) and TOSCA (Tracking Oil Spills and Coastal Awareness, G-MED09-425) both funded by the MED-INTEREG IVA program from the European Commission and project MIDAS-7 (AYA2012-39356-C05-03) funded by the Spanish National Research program funded by MICIN. José Antonio Jiménez Madrid is under contract by MEDESS-4MS.

The oceanic daily outputs have been provided by the SiMED french project funded by GMMC. The thank ENSTA-ParisTech and Mercator Ocean for providing the oceanic fields. The NEMO-MED12 simulation was granted access to the HPC resources of IDRIS of CNRS (project number 010227) made by GENCI. We thank M. Déqué from Météo-France for running the ARPERA simulation.

\bibliography{risk_mpb}

\clearpage
\section{Figures}

\begin{figure}[htb]
\figura[width=\linewidth]{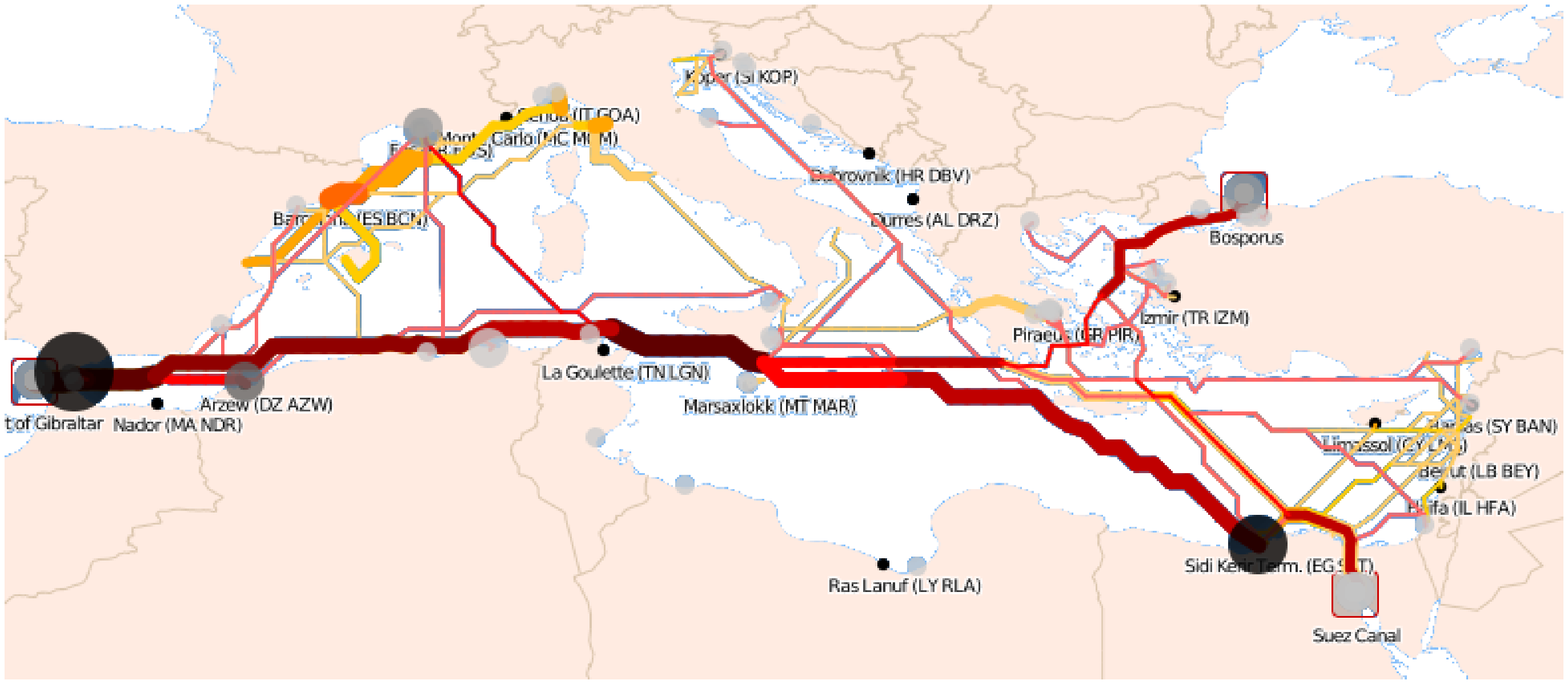}
\caption{Main paths of oil tankers through the Mediterranean Sea. Black and grey dots are main ports were crude oil tankers have been calling in year 2005. Red lines are main routes for crude oil tankers for year 2005 and lines in orange are main traffic routes for the same year. (Source: Med GIS on Maritime Traffic from EU funded Safemed Project, see http://www.safemedgis.org/)}
\end{figure}

\begin{figure}[htb]
\figura[width=\linewidth]{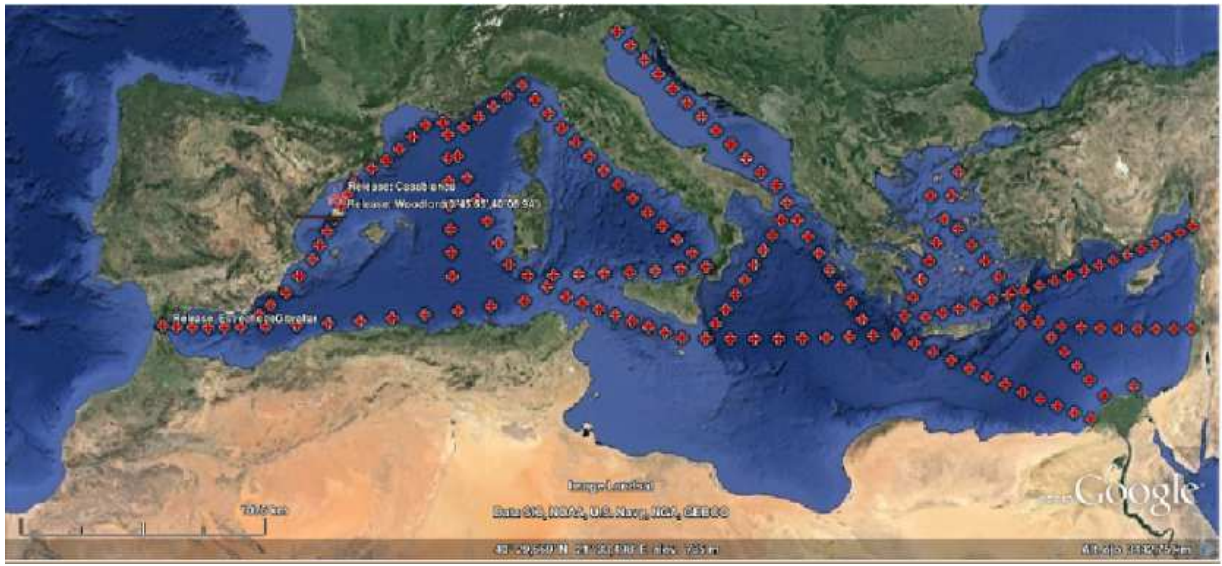}
\caption{Selected points along main paths of oil tankers, including some
interesting points like Casablanca platform, Woodford vessel and some points
selected in the MED-GIB experiment.\label{fig:spillpoints}}
\end{figure}

\begin{figure}[htb]
\figura[width=\linewidth]{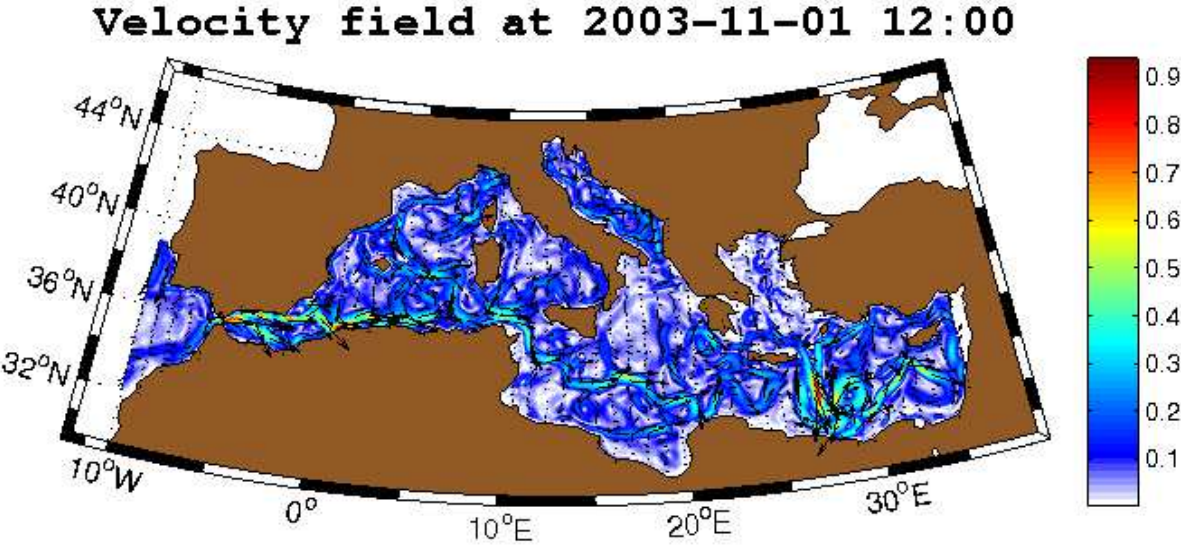}
\caption{Snapshot of a velocity field hindcast produced by NEMO-MED12.\label{fig:velocityfield}}
\end{figure}

\begin{figure}[htb]
\figura[width=\linewidth]{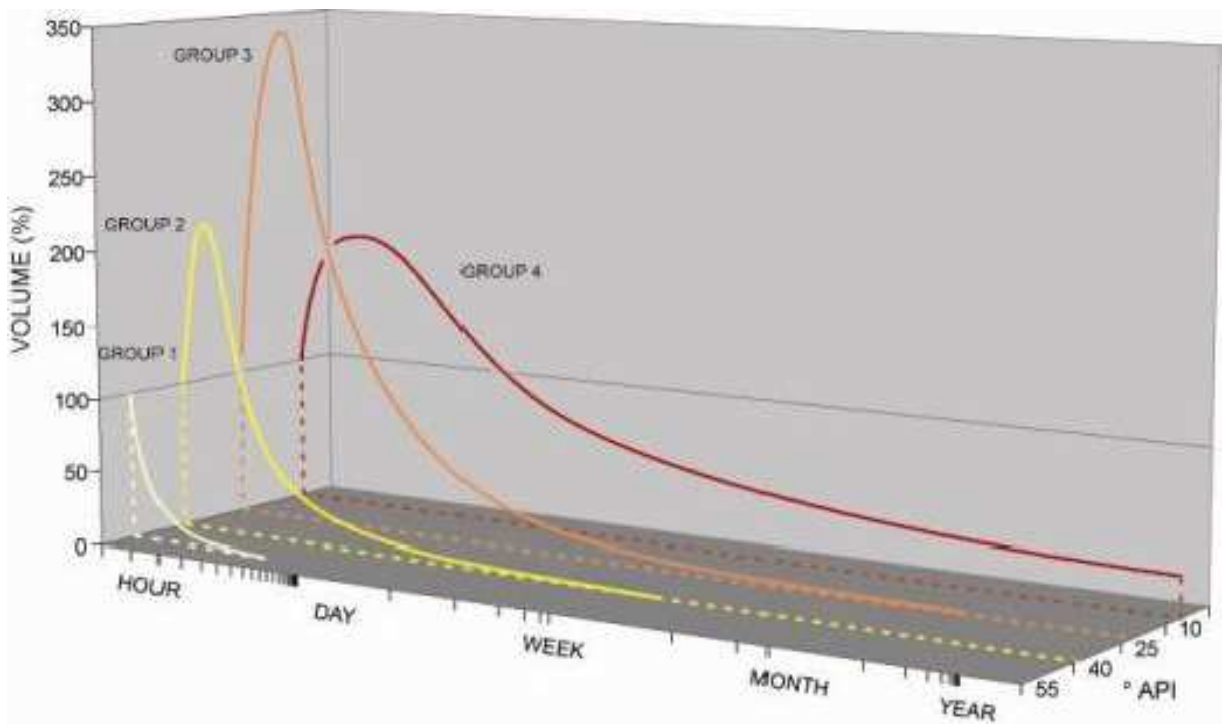}
\caption{Oil persistence curves according to the ITOPF documentation. Adapted from http://www.itopf.com/marine-spills/fate/models/index.html.
\label{fig:persistence}}
\end{figure}

\begin{figure}[htb]
\figura[width=\linewidth]{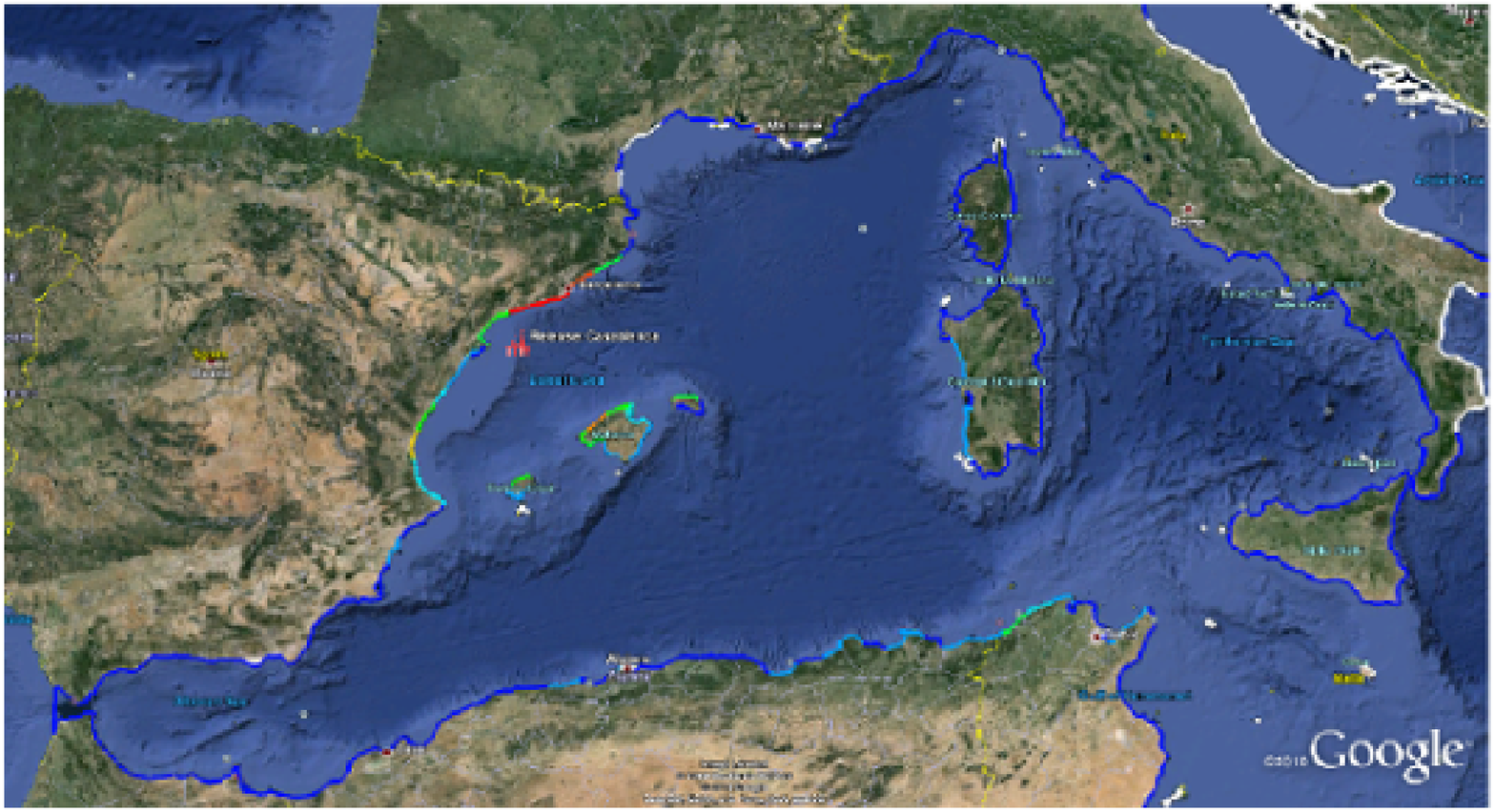}
\figura[scale=0.70,bb = 15 -282 115 -182]{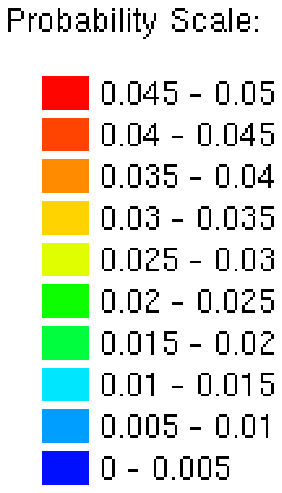}
\caption{Beaching probability map for a spill produced at the Casablanca oil platform (the red platform icon in the image). \label{fig:casablanca}}

\end{figure}

\begin{figure}[htb]
\figura[width=\linewidth]{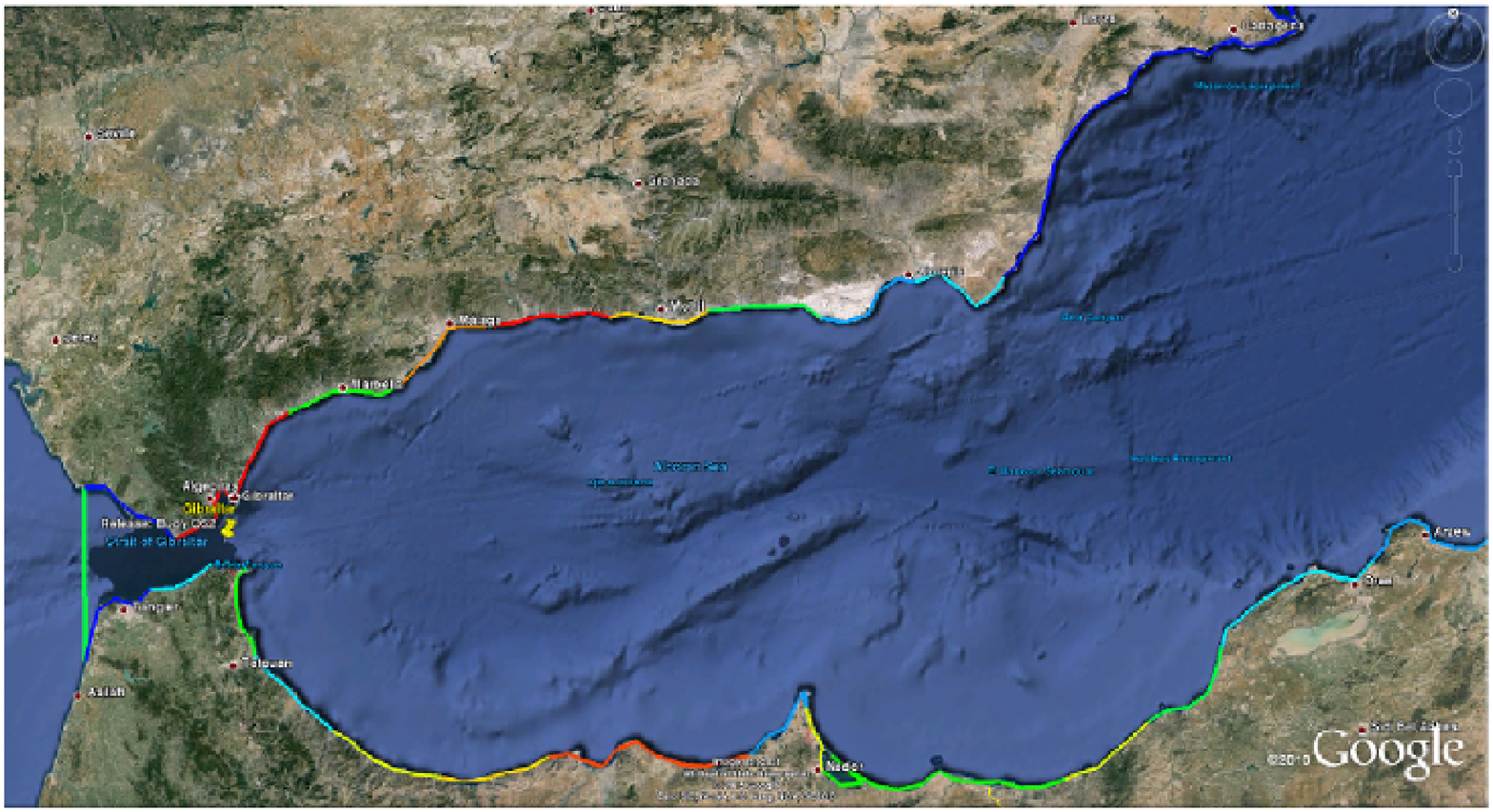}
\figura[width=\linewidth]{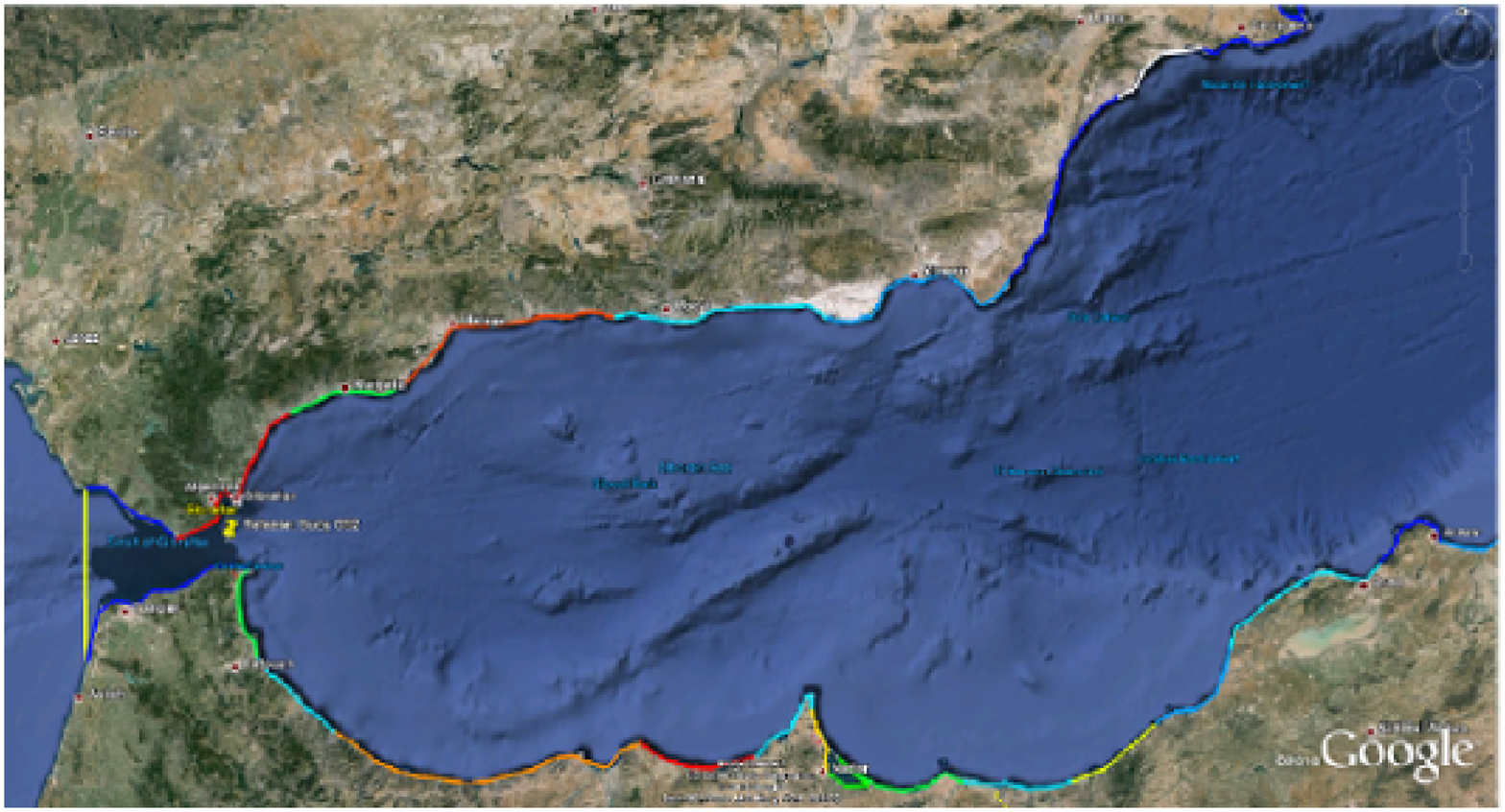}
\figura[scale=0.70,bb = 15 -632 115 -532]{scale}
\caption{Beaching probability maps corresponding to a source point in the Gibraltar Strait. Top panel: Map considering the full period of hindcast data. Bottom: Map considering particles released only during the first fortnight of every September.\label{fig:gibraltar}}
\end{figure}

\begin{figure}[htb]
\begin{center}
\figura[width=0.49\linewidth]{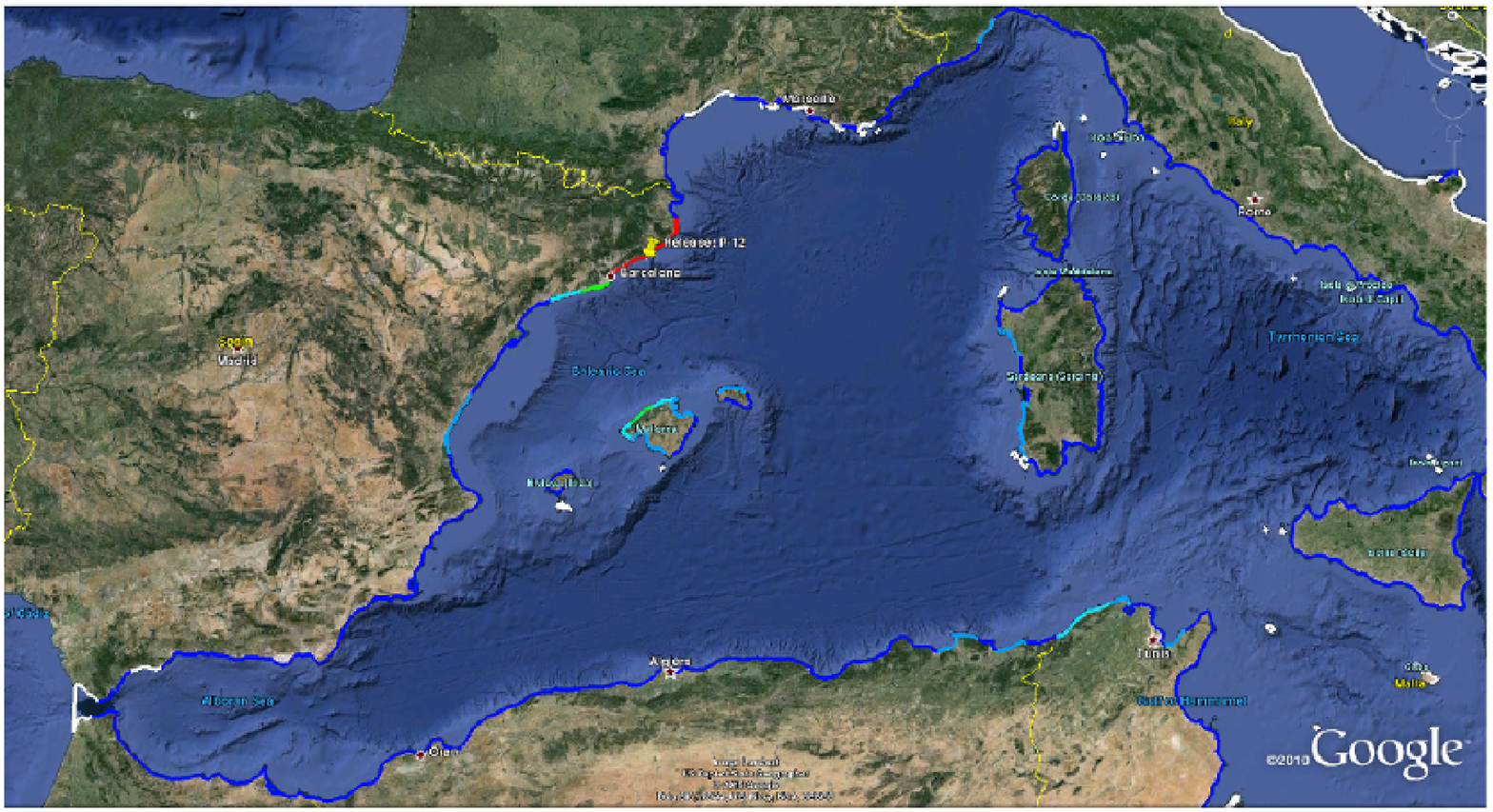}
\figura[width=0.49\linewidth]{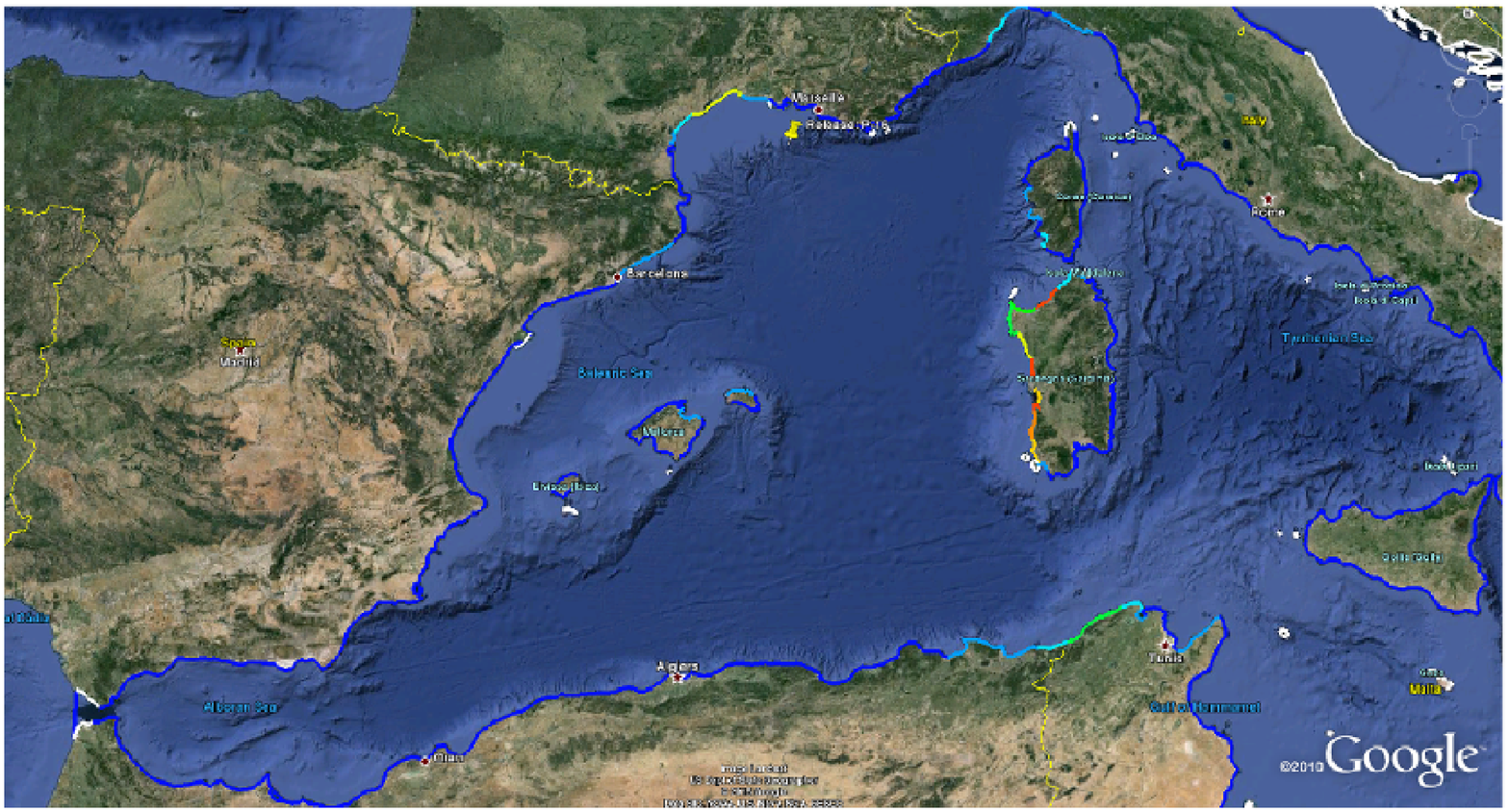}
\figura[width=0.49\linewidth]{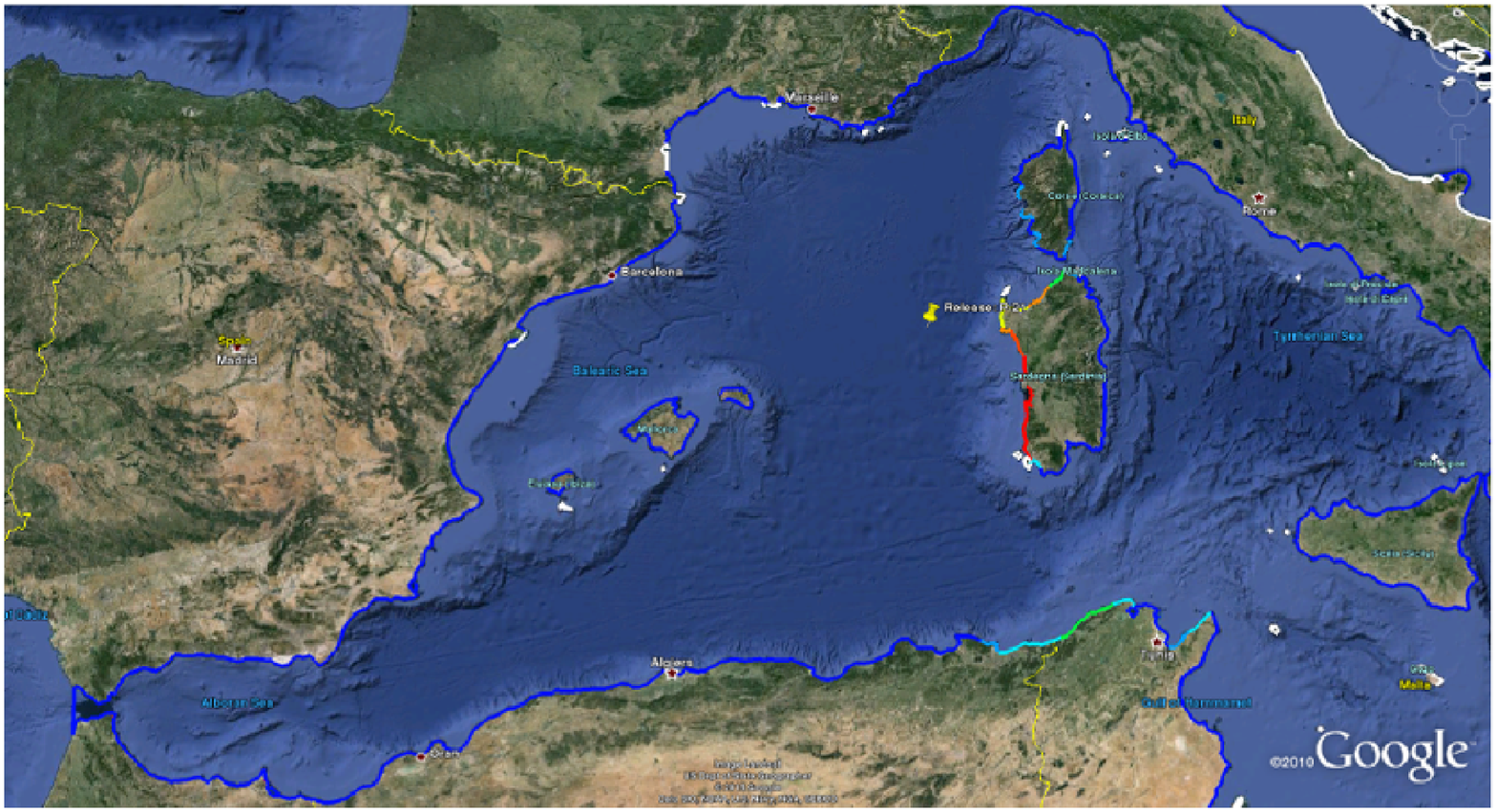}
\figura[width=0.49\linewidth]{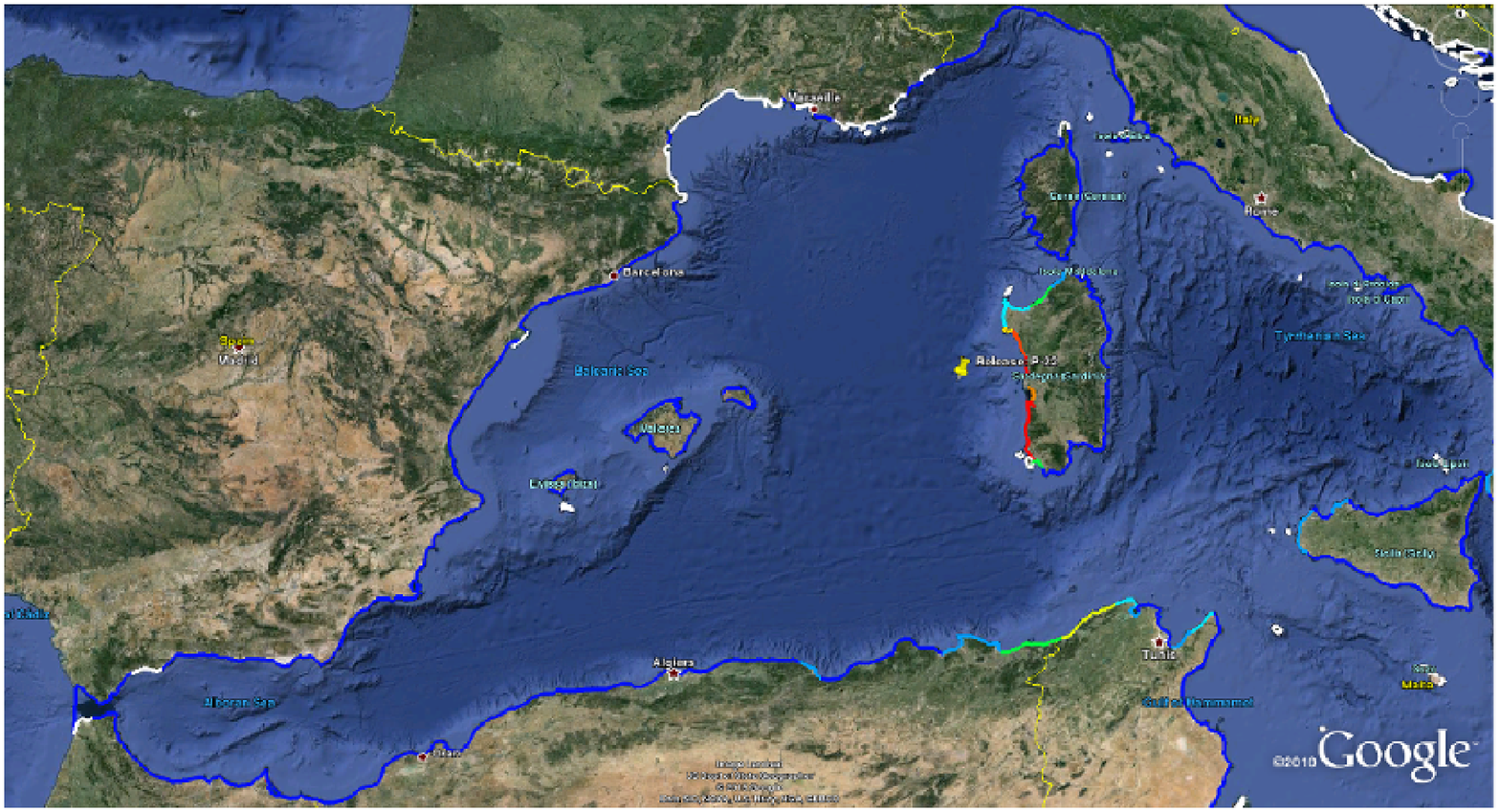}
\end{center}
\caption{Beaching probability maps corresponding to release points labelled as 12 (top-left), 16 (top-right), 21 (bottom-left)  and 22 (bottom-right) respectively. The points are indicated in the plot by the yellow drawing pin. The colour scale is the same than in the previous figure.\label{fig:barrier}}
\end{figure}


\begin{figure}[htb]
\begin{center}
\figura[width=0.75\linewidth]{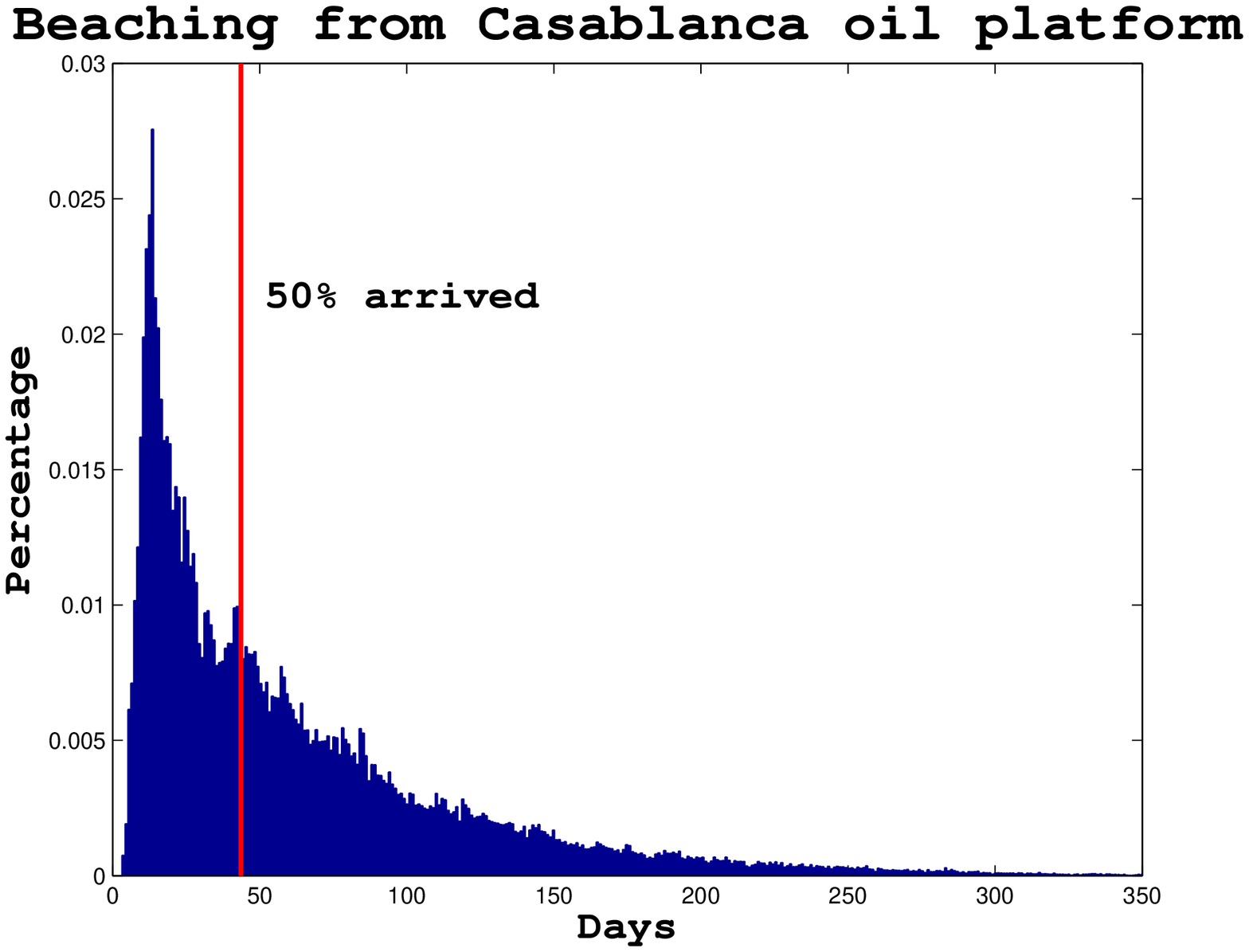}
\figura[width=0.75\linewidth]{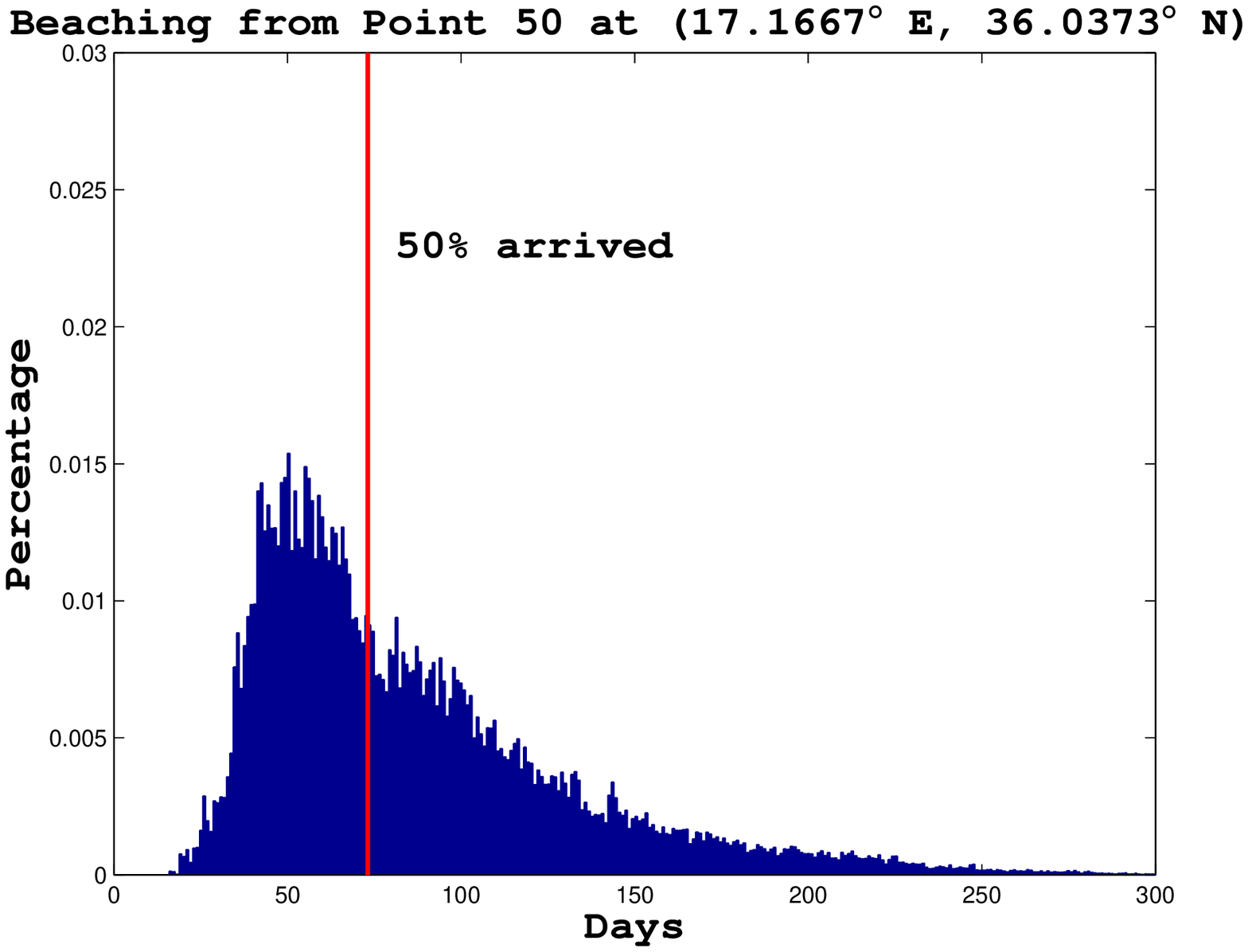}
\end{center}
\caption{Percentage of particles reaching the coast as function of
time required to arrive. At the top is depicted the time needed
for an oil spill from the Casablanca oil platform and in the bottom 
coming from a point located at (17.1667º E, 36.0373º N). The vertical
red line indicates the time when the $50\%$ of particles have arrived to
one coast. 
 \label{fig:percentages}}
\end{figure}
\end{document}